\documentclass[aps,twocolumn,superscriptaddress,floatfix]{revtex4-1}

\usepackage{lipsum}
\usepackage{graphicx}
\usepackage{amsmath,amssymb}
\usepackage{braket}
\usepackage{mathtools}
\usepackage{hyperref}
\usepackage[version=4]{mhchem}
\usepackage{siunitx}

\usepackage{color}
\usepackage[normalem]{ulem}

\usepackage{xcolor}

\newcommand{\Z}{\mathbb{Z}}

\renewcommand{\v}[1]{\boldsymbol{#1}}

\newcommand{\milli}{{\rm m}}
\newcommand{\mili}{{\rm m}}
\newcommand{\electronvolt}{{\rm eV}}
\renewcommand{\S}{\widehat{\mathbf{S}}}

\DeclareSIUnit\muBFe{\micro_B \per Fe}

\begin{document}

\title{Half-magnetization plateau and the origin of threefold symmetry breaking in an electrically-switchable triangular antiferromagnet}

\newcommand{\UCB}{Department of Physics, University of California, Berkeley, CA 94720, USA}
\newcommand{\LBL}{Materials Sciences Division, Lawrence Berkeley National Laboratory, Berkeley, California, 94720, USA}
\newcommand{\maglab}{National High Magnetic Field Laboratory, Tallahassee, Florida 32310, USA}
\newcommand{\kavli}{Kavli Energy Nanosciences Institute at Berkeley, Berkeley, CA 94720}

\author{Shannon C. Haley}
\affiliation{\UCB}
\affiliation{\LBL}

\author{Sophie F. Weber}
\affiliation{\UCB}
\affiliation{\LBL}

\author{Taylor Cookmeyer}
\affiliation{\UCB}
\affiliation{\LBL}

\author{Daniel E. Parker}
\affiliation{\UCB}

\author{Eran Maniv}
\affiliation{\UCB}
\affiliation{\LBL}

\author{Nikola Maksimovic}
\affiliation{\UCB}
\affiliation{\LBL}

\author{Caolan John}
\affiliation{\UCB}

\author{Spencer Doyle}
\affiliation{\UCB}

\author{Ariel Maniv}
\affiliation{\maglab}
\affiliation{NRCN, P.O. Box 9001, Beer Sheva, 84190, Israel}

\author{Sanath K. Ramakrishna}
\affiliation{\maglab}

\author{Arneil P. Reyes}
\affiliation{\maglab}

\author{John Singleton}
\affiliation{National High Magnetic Field Lab (NHMFL), Los Alamos National Lab (LANL), Los Alamos, NM 87545, USA}

\author{Joel E. Moore}
\affiliation{\UCB}
\affiliation{\LBL}

\author{Jeffrey B. Neaton}
\affiliation{\UCB}
\affiliation{\LBL}
\affiliation{\kavli}

\author{James Analytis}
\email[]{shannon\_haley@berkeley.edu or {analytis@berkeley.edu}}
\affiliation{\UCB}
\affiliation{\LBL}

\date{\today}

\begin{abstract}
	We perform high-field magnetization measurements on the triangular lattice antiferromagnet \ce{Fe_{1/3}NbS_2}. We observe a plateau in the magnetization centered at approximately half the saturation magnetization over a wide range of temperature and magnetic field. From density functional theory calculations, we determine a likely set of magnetic exchange constants. Incorporating these constants into a minimal Hamiltonian  model of our material, we find that the plateau and of the $\Z_3$ symmetry breaking ground state both arise from interplane and intraplane antiferromagnetic interactions acting in competition. These findings are pertinent to the magneto-electric properties of \ce{Fe_{1/3}NbS_2}, which allow electrical switching of antiferromagnetic textures at relatively low current densities.

\end{abstract}

\maketitle

The electrical manipulation of antiferromagnetic spin textures has the potential to effect transformative technological change \cite{jungwirth_multiple_2018}. Exotic magnets with complex interactions are of special interest in this field, because they are likely to leverage novel mechanisms for their manipulation, possibly allowing ultra low-power or ultra-fast functionality. Diagnosing the relative magnitude of these interactions gives a direct insight into these mechanisms. The existence of magnetization plateaus at fractions of saturation, when a material is subjected to large external magnetic fields, is a powerful tool to this end~\cite{Takigawa2011}.

In this work we study magnetization plateaus in the antiferromagnet \ce{Fe_{1/3}NbS_2}, a magnetically intercalated transition metal dichalcogenide which has recently been found to exhibit reversible, electrically-stimulated switching between stable magnetic states~\cite{nair_electrical_2019}. This behavior has been seen with considerably lower energy requirements in \ce{Fe_{1/3}NbS_2} as compared to the other systems~\cite{nair_electrical_2019}, raising the question of whether the mechanism differs significantly~\cite{zelezny_relativistic_2014,wadley_electrical_2016}. At the center of this question is the nature of the magnetic ground state, which has been challenging to determine because collinear and non-collinear order are energetically close and the true ground state depends strongly on the magneto-crystalline anisotropy ~\cite{mankovsky_electronic_2016}. The nature of the underlying ordering in \ce{Fe_{1/3}NbS_2} has been studied by both neutron scattering~\cite{van_laar_magnetic_1971,Suzuki1993} of magnetic order and optical linear birefringence microscopy~\cite{little_observation_2019}, which probes nematic structure in the electrical conductivity. Both measurements --- electric and magnetic --- find indications of three-fold symmetry breaking in the ground state, whose origin is unclear.

We report here a hitherto unobserved plateau in the field-induced magnetization at half of the saturation value. Such a plateau has previously been argued to exist theoretically in both anisotropic classical~\cite{seabra2011competition} and isotropic quantum~\cite{ye2017half} models of triangular lattice antiferromagnets.  In the isotropic case, the half-magnetization plateau exists whenever there is a significant next nearest neighbor magnetic coupling~\cite{PhysRevB.95.014425}. (We will discuss the magnetic Hamiltonian inferred from magnetization measurements and spin-wave analysis in more detail below.) Plateaus at one third of the saturation magnetization have been studied extensively in triangular systems; their occurrence is often explained by the stabilization of an up-up-down state by quantum fluctuations (part of a phenomenon known as order-by-disorder) \cite{PhysRevB.67.104431, PhysRevB.67.094434,WIEDENMANN19887,doi:10.1143/JPSJ.80.093702,Gvozdikova_2011,PhysRevLett.110.267201,PhysRevLett.102.137201}. However, experimental realizations of a half-magnetization plateau on a triangular lattice are relatively rare~\cite{coldea_cascade_2014,WIEDENMANN19887}, and their observation is strong evidence for the presence of significant interactions beyond the first nearest neighbor. 

The implication from theory is that the same interactions that generate the plateau are also responsible for a threefold symmetry breaking stripe phase in the ground state, for both quantum and classical models.  The half-magnetization plateau found in \ce{Fe_{1/3}NbS_2} thus gives a strong clue to the physical mechanism of the threefold symmetry breaking in this material in zero applied field, and thereby provides a microscopic picture for the electrically switchable antiferromagnetic texture.

\begin{figure}[t]
    \centering
    \includegraphics[width=\linewidth]{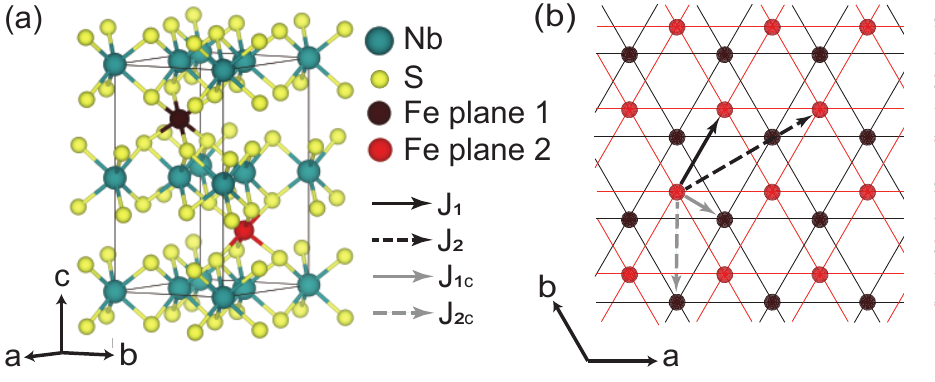}
    \caption{(a) The crystal structure of \ce{Fe_{1/3}NbS2}. Iron atoms sit between layers of \ce{NbS2}, aligned with the niobium atoms above and below. (b) Looking along the c-axis, the iron atoms in a given layer form a triangular lattice. These triangular lattices are shifted from layer to layer. Arrows indicate in-plane and out-of-plane first and second nearest neighbors, labeled by their relevant exchange constants.
    }
    \label{fig:structure}
\end{figure}

\ce{Fe_{1/3}NbS_2} is a layered material with space group $P6_322$ {\textbf{\textit No. {182}}} whose magnetism arises from the iron which sits between layers of \ce{NbS2} (Fig.~\ref{fig:structure} (a)). These magnetic atoms form triangular lattices in each layer, with adjacent layers staggered with respect to one another (Fig.~\ref{fig:structure} (b)). Charge from the iron atoms is transferred to the NbS$_2$ conduction band, leaving them in a 2+ ionized state, with four unpaired localized electrons per atom~\cite{cw4p3_2plus,mossbauer}. The macroscopic behavior of the material in low field is antiferromagnetic (AFM).
The samples discussed in this study were grown via chemical vapor transport, as described in Ref. \cite{doyle2019tunable}. Using EDX and ICP, the ratio of Fe:Nb was found to be 0.330:1.

\begin{figure}[t]
    \centering
    \includegraphics[width=\linewidth]{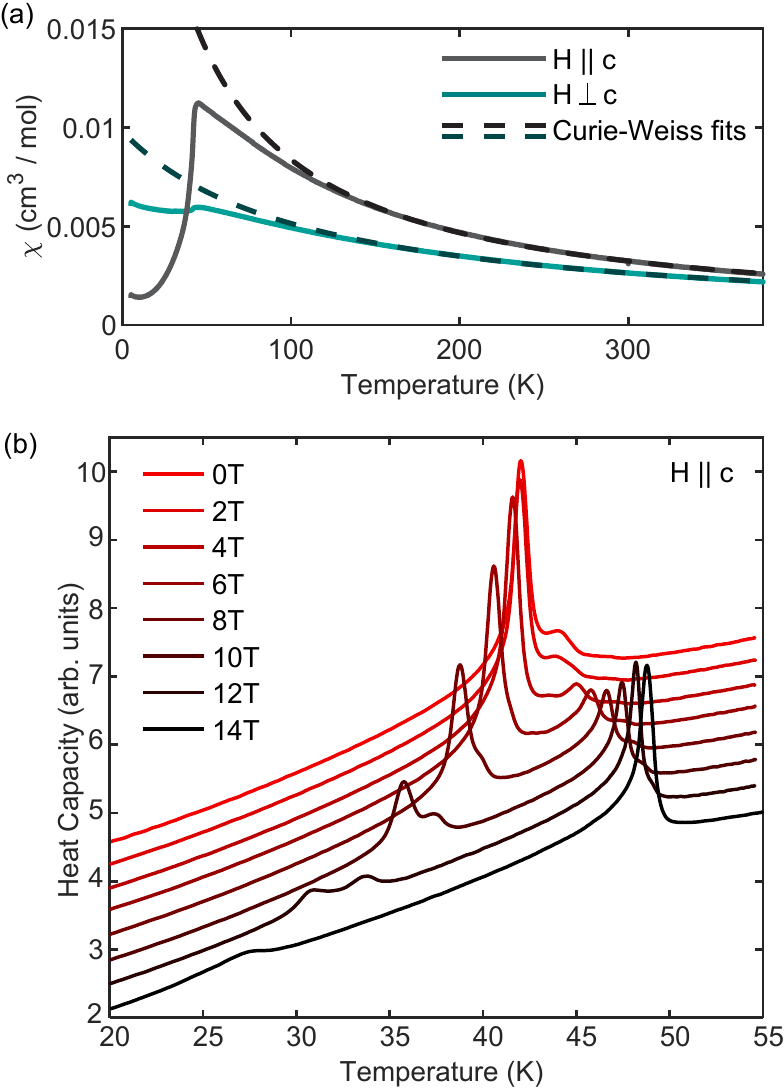}
    \caption{(a) Curie-Weiss fits of both out of plane ($H\parallel c$) and in plane ($H\perp c$) susceptibility yield an effective moment of $\mu_{\text{eff}} = 5$~\si{\muBFe}, and single-ion anisotropy of approximately $D = $~\SI{1}{\milli\electronvolt}. (b) Heat capacity measurements show two transitions, which split with the application of field parallel to the c-axis. Curves are offset to enhance visibility. }
    \label{fig:cw}
\end{figure}

Measurements of the magnetic susceptibility as a function of temperature in low applied fields show AFM behavior below a transition near 45K (Fig.~\ref{fig:cw} (a)). Fitting to the paramagnetic regime, the Curie-Weiss Law yields an estimate of \SI{ 5 }{\muBFe} for the effective moment of the material. This is in agreement with the values found in the literature, which predominantly range from \SIrange[range-units = single]{4.3}{5}{\muBFe}~\cite{cw4p3_2plus,cw4p44,cw4p94,cw4p6,cw5ish}, although there is one report as high as 6.3\si{\muBFe}~\cite{friend}. It should be noted that this is thought to include a significant orbital contribution~\cite{friend}. In general, the effective moment is additionally distinct from and slightly higher than the expected moment at saturation. 

Heat capacity measurements resolve two clear transitions at zero field (Fig.~\ref{fig:cw} (b)). With the application of field, these transitions move apart from each other in temperature. The lower temperature transition has a further splitting at higher fields, indicating the presence of an additional intermediate phase.

\begin{figure}
    \centering
    \includegraphics[width=\linewidth]{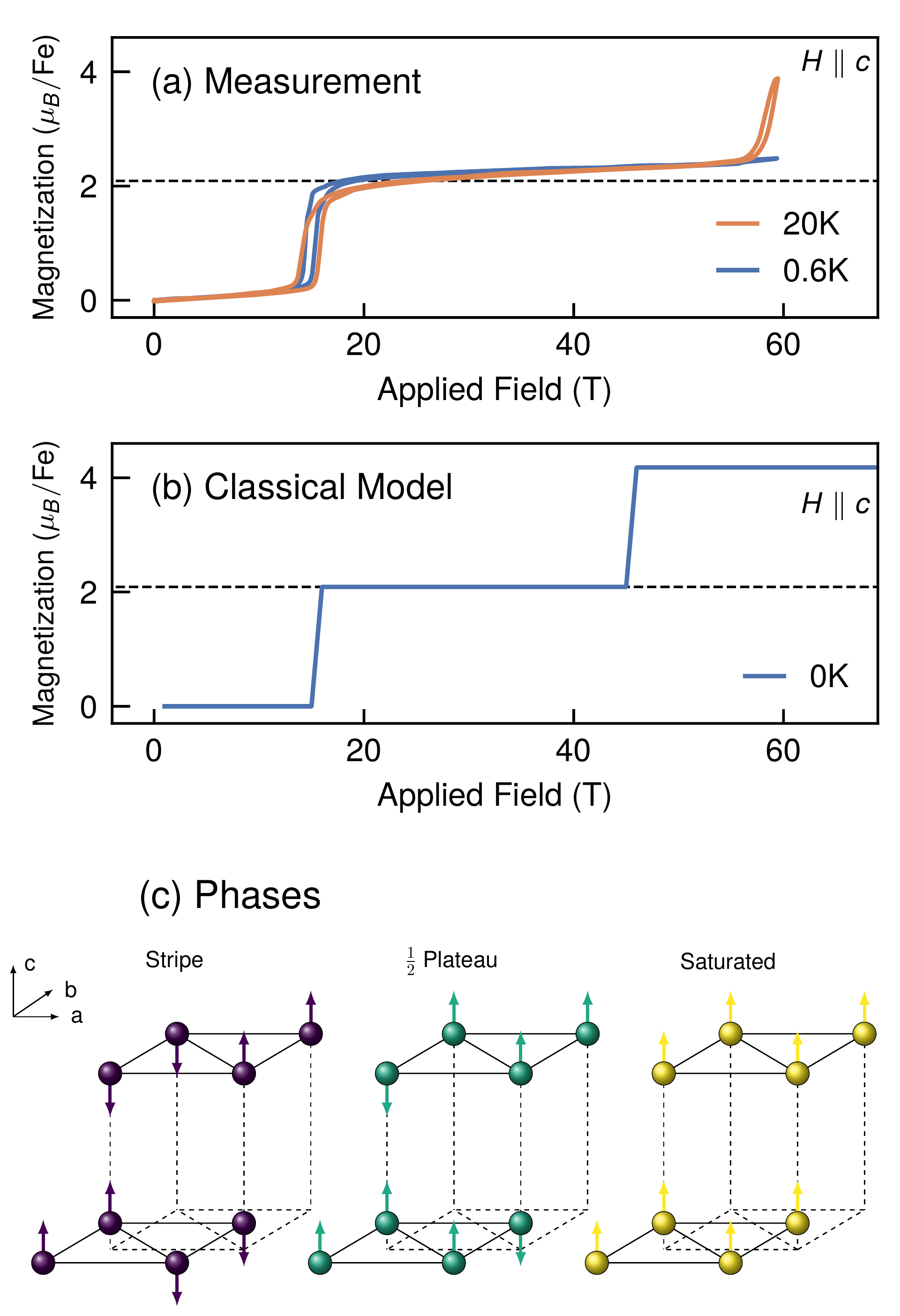}
    \caption{(a) Magnetization response of \ce{Fe_{1/3}NbS_2} to an out-of-plane pulsed field. (Data from a 25T pulse is used below 15T for the 0.6K curve.) At \SI{0.6}{\kelvin}, the magnetization shows two flat plateaus at $0$ and $1/2$ of the saturated magnetization (dashed line). At \SI{20}{\kelvin} a further transition, likely to a fully saturated state, is observed near \SI{60}{\tesla}. (b) Magnetization response of the model, Eq. \eqref{eq:mod_ham}, computed classically. Three plateaus are clearly visible: a stripy AFM phase, a UUUD phase, and a saturated PM phase. Calculational details are given in the Supplemental Materials~\cite{SupMat}. (c) Cartoons of the spin configurations in the eight site unit cell.
    }
    \label{fig:magnetization}
\end{figure}

\begin{figure}[t]
    \centering
    \includegraphics[width=\linewidth]{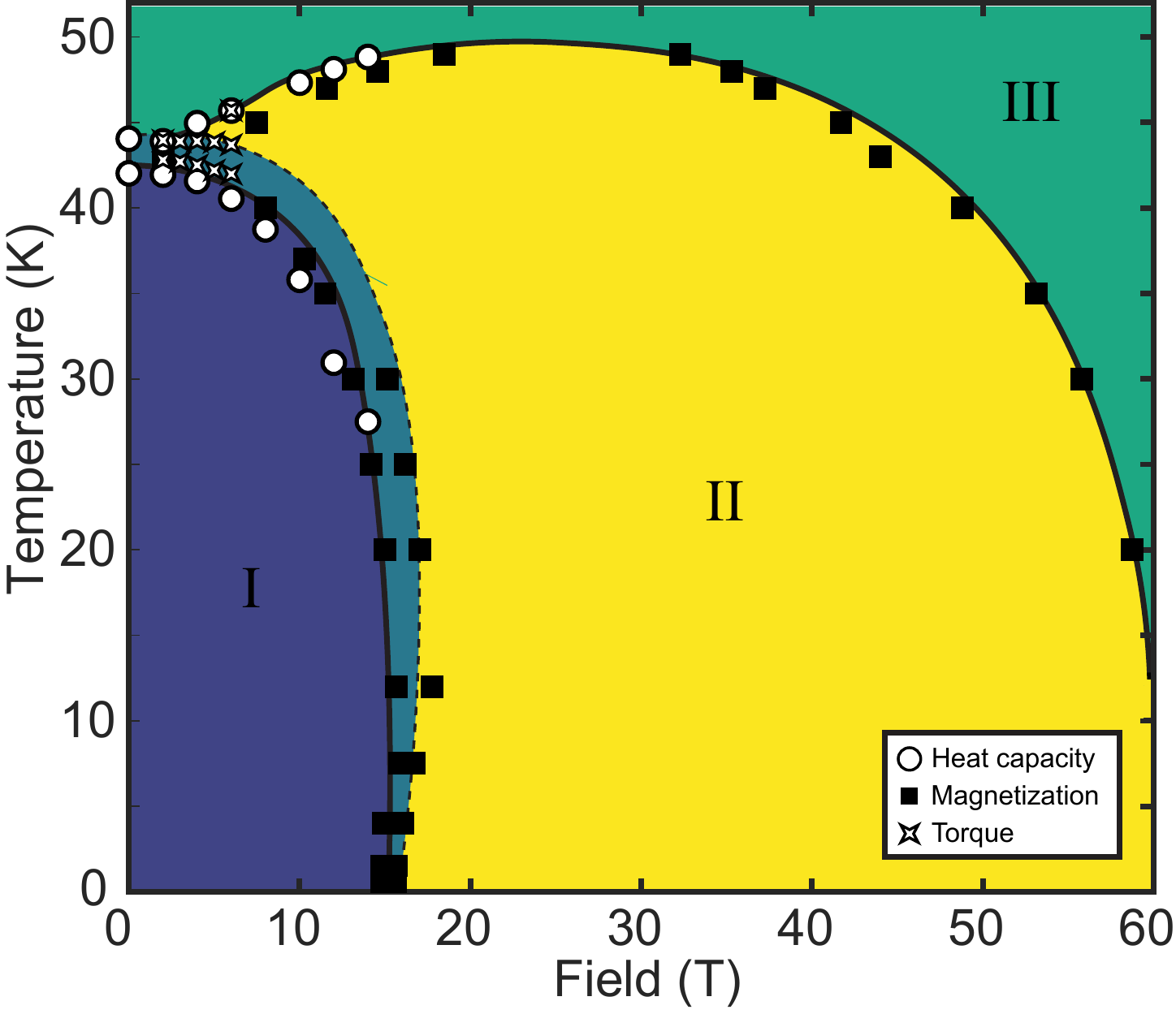}
    \caption{The full experimental phase diagram of \ce{Fe_{1/3}NbS2}, as a function of temperature and field applied along the c-axis. The two phases visible at low temperature are different antiferromagnetic orderings, corresponding to (nearly) zero net magnetization and the 1/2 magnetization plateau (I and II respectively). Our calculations suggest that at low temperature, the former is a stripe phase while the plateau is UUUD. An intermediate phase bridging the stripe and UUUD phase has also been found, of unknown origin. This phase boundary was determined in part via torque magnetometry; more details on this measurement are found in \cite{SupMat}. Good agreement is seen between phase boundaries determined by the heat capacity and pulsed field magnetization measurements. Phase boundary lines are a guide to the eye.}
    \label{fig:phasediagHT}
\end{figure}

High field measurements further elucidate the nature of the phase transitions. Measurements at 0.6\si{\kelvin} and 20\si{\kelvin} of the magnetization as a function of applied field are shown in Fig.~\ref{fig:magnetization} (a). The full set of measurements, taken at temperatures ranging from 0.6\si{\kelvin} to 50\si{\kelvin}, is given in \cite{SupMat}, and the phase boundaries determined in part from these measurements are shown in Fig.~\ref{fig:phasediagHT}. These measurements were performed on a stack of about 30 co-aligned crystals, which were roughly 1mm in diameter and had an average thickness of 0.1mm.

From these measurements it can be seen that there are three dominant phases at low temperature: (I) the zero field phase characterized by a small magnetic moment, (II) the `plateau' phase characterized by a nearly constant magnetic moment centered around half the estimated saturation moment, and (III) a high field phase which approaches the fully saturated moment. The final phase gets pushed above 60T at the lowest temperatures. An intermediate phase bridging the zero field and plateau phase has only a weak feature in the magnetization (see Supplemental Material \cite{SupMat}).

The experimental phase diagram, Fig.~\ref{fig:phasediagHT}, shows a nonmonotonic dependence of the ordering temperature on applied field. This dependence can be explained by the impact of an applied field on a reduced dimensional system, as was argued by Sengupta {\it et al.} regarding the tetragonal AFM [Cu(HF$_2$)(pyz)$_2$]BF$_4$~\cite{sengupta_nonmonotonic_2009}. Fluctuations of the mean field become significant in such systems. As the field increases, both the order parameter and these fluctuations are suppressed. The latter effect leads to an increase in the transition temperature in low field, and the former takes over and brings down the transition temperature at higher fields. In our case, there is the interesting addition of a second ordered phase, which is destroyed in that low field regime.

While the primary measurements were performed on a stack of co-aligned crystals in pulsed field, the nature of the plateau was confirmed both with a single crystal in pulsed field, and with a stack of crystals in a \SI{30}{\tesla} DC field. A comparison of these measurements to pulsed field is found in the Supplement \cite{SupMat}. The DC measurement was additionally used to scale the pulsed field data, whose experimental apparatus measures magnetization up to a constant scaling factor. The effect of the slight deviation from the perfect 1:3 Fe:Nb ratio was examined by measuring another growth with an Fe:Nb ratio of 0.339:1, to be compared to our primary growth with a ratio of 0.330:1. The underlying behavior is largely unchanged, although there is some movement of the phase boundaries. This data can be found in the Supplement \cite{SupMat}.

To understand the physical mechanism responsible for the magnetization plateaus (Fig.~\ref{fig:magnetization}(a)), we study a minimal model motivated by our density functional theory (DFT) calculations, the details of which will be discussed later in the text. In addition to the single-ion anisotropy \ce{D}, we find that a model with nearest neighbor (NN) and next nearest neighbor (NNN) exchange couplings within a single \ce{Fe} plane, as well as NN and NNN couplings between adjacent planes, is sufficient to accurately reproduce the ab-initio energies of various magnetic states.
We restrict our attention to the \ce{Fe} atoms and their localized $d$ states, which form a lattice of $S=2$ spins, and consider a short-range Hamiltonian 
\begin{align}
	\label{eq:mod_ham}
	\widehat{H} & = E_0 + 2J_1 \sum_{\braket{i,j}} \S_i \cdot \S_j + 
	2J_2 \sum_{\braket{\braket{i,j}}} \S_i \cdot \S_j \\ 	\notag
	&+ 2J_{1c} \sum_{\braket{i_c,j_c}} \S_i \cdot \S_j + 2J_{2c} \sum_{\braket{\braket{i_c,j_c}}} \S_i \cdot \S_j
	- \sum_i D \left( \widehat{S}_i^z \right)^2,
\end{align}
where $J_1$ and $J_2$ are the nearest-neighbor (NN) and next-nearest-neighbor (NNN) exchange couplings within a single \ce{Fe} plane, $J_{1c}$ and $J_{2c}$ are the NN and NNN couplings between adjacent planes and, crucially, $D$ is the magnetoanisotropy of \ce{Fe} spins.
 ~$E_{0}$ encompasses any non-magnetic contributions to the total energy.
The exchange coupling sums are over all unique bonds. 
We will see that, in a large neighborhood of relevant exchange coupling values, this model has three distinct phases at zero temperature as the magnetic field is varied. (1) An ``AFM stripe'' phase at low field with a magnetic unit cell of 4 \ce{Fe} spins, with 2 pointing up along $+c$ and 2 along $-c$ in a stripe configuration. (2) A half-magnetization plateau at intermediate field with a magnetic unit cell of $8$ \ce{Fe} spins, with three up spins and one down spin per layer (denoted UUUD). (3) A saturated phase at high field with a magnetic unit cell of $2$ \ce{Fe} spins which are all pinned to point up, parallel to $\v{H}$. These configurations are shown in Fig. \ref{fig:magnetization} (c).

 Half-magnetization plateaus have been observed before in two close antecedents of Eq. \eqref{eq:mod_ham} in the literature --- one with $D = J_{1c}=J_{2c}= 0$ in \cite{ye2017half} and another with different stacking of the triangular layers \cite{seabra2011competition}. In reference \cite{ye2017half}, a ``stripy'' AFM phase is preferred at low fields, and there is a transition to a UUUD phase, which is stabilized by the ``order-by-disorder'' mechanism. In reference \cite{seabra2011competition}, the introduction of an interplanar interaction, and $D\ne 0$ broadens the region of stability for the UUUD phase at the classical level, so it makes sense to expect plateaus in Eq. \eqref{eq:mod_ham}.

Due to the spins being large ($S=2$), we perform a classical analysis of Eq.~\eqref{eq:mod_ham}. We search for the ground state of Eq.~\eqref{eq:mod_ham} using many different sized trial unit cells. While a fully 3D classical Monte-Carlo simulations would be more exhaustive, the present analysis is sufficient because high-field measurements of the nuclear magnetic resonance suggest that the plateau has a relatively simple spin texture \cite{SupMat}. More details can be found in the Supplement \cite{SupMat}. We find that the magnetic unit cell for the ground state is always small over a very broad range of parameters $J$, $D$, and $h$, with no more than 8 \ce{Fe} atoms. Intuitively, this small unit cell is consistent with the short-ranged nature of the dominant interactions. 

The classical analysis shows there is a large range of couplings ($J_1,J_2,J_{1c}, J_{2c})$ which produce the three phases observed as a function of magnetic field when $D>0$ is large. The key observation is that, for $J_1 > 0$ and $J_2/J_1 \ll 1$, there is a large region in the $(J_{1c}, J_{2c})$ parameter space that approximately reproduces the magnetization curves -  the ``stripy'' AFM, UUUD, and UUUU are the only three ground states for a wide range of $J_{1c}/J_1 >-1$ and $J_{2c}/J_1 <0$. In fact, the only 1/2-magnetization plateau without a UUUD structure between the two layers occurs for only a small region of parameter space. Phase diagrams are given in the Supplement \cite{SupMat}. We may conclude that Eq. \eqref{eq:mod_ham} qualitatively reproduces the observed transitions in the magnetization even without precise estimates for the coupling parameters.

We now quantitatively predict the critical magnetic fields for the transitions from the model Eq. \eqref{eq:mod_ham}. For large $D>0$, the transition from the stripe phase to the plateau phase occurs when $h = 4(J_1+J_{1c}+J_2)$ and the transition from the plateau phase to the saturated phase occurs when $h=12(J_1+J_{1c}+J_2)$. Quantitative analysis requires estimates of the parameters $(J_1,J_2,J_{1c},J_{2c},D)$, which we now ascertain through a combination of experimental and numerical means. Following Ref. \cite{DfromCW}, we can relate the magneto-crystalline anisotropy $D$ to the in- and out-of-plane Curie-Weiss temperatures, which are found from the fits in Fig. \ref{fig:cw}(a) to be \SI{-110}{\kelvin} and \SI{-26}{\kelvin}, respectively; this analysis yields $D\approx  \SI{1}{\milli\electronvolt}$. While Ref. \cite{cw4p3_2plus} gives slightly lower Curie-Weiss temperatures (\SI{-135}{\kelvin} and \SI{-40}{\kelvin}), these values give a virtually unchanged estimate of $D$, which is proportional to their difference.

Our DFT calculations, performed with the Perdew-Burke-Ernzerhof (PBE) functional \cite{Perdew1996} and Hubbard U corrections \cite{Perdew1986}, corroborate this picture. We note that the calculated $D$, being a highly local property, is sensitive to the Hubbard $U$ used to approximately treat the localized \ce{Fe} $d$ electrons. This sensitivity has been documented for several \ce{Fe}-based compounds in previous literature \cite{Yang2001,Bousquet2010}. However, the experimental estimate of $D$ allows us to choose a $U$ value that yields a similar anisotropy, and with which to compute the exchange constants in the  minimal model. Using a Hubbard $U$ of $\SI{0.3}{\electronvolt}$ in our PBE+U calculations at experimental lattice parameters (see supplement \cite{SupMat} for details), we obtain $D=\SI{1.09}{\milli\electronvolt}$, with the easy axis along \ce{c}, in very good agreement with experiment. 

We then compute the Heisenberg exchange parameters with PBE+U and $U=\SI{0.3}{\electronvolt}$. Using six inequivalent magnetic collinear configurations with \ce{Fe} spins along the $c$ axis, we solve an overdetermined system of equations for the unknown couplings (note that the anisotropy cancels out since all configurations have spins aligned in the same direction). The values of all $J$ as well as $D$ are given in Table \ref{tab:couplings}.
\begin{table}[h!]
\centering
\begin{tabular}{ccccc} 
 \toprule
$D$ &  $J_1$ & $J_2$ & $J_{1c}$ & $J_{2c}$ \\ 
 \hline
$1.09$ & 0.76 & -0.006 & 0.39 & -0.22 \\ 
\botrule
\end{tabular}
\caption{PBE+U ($U=\SI{0.3}{\electronvolt}$) values of magneto-crystalline anisotropy $D$ and NN and NNN interplanar and intraplanar couplings in Eq. \ref{eq:mod_ham}. Units are \si{\mili\electronvolt} per \ce{Fe} atom. With the conventions used in Eq. \ref{eq:mod_ham}
positive values for $J$ represent AFM couplings, negative values are FM, and a positive value of $D$ implies an easy-axis along
c for the anisotropy.}
\label{tab:couplings}
\end{table}

As an experimental check, the Curie-Weiss temperatures can be related to the sum of all coupling constants, giving an estimate $\sum_i J_i = 6(J_1 + J_2 + J_{1c} + J_{2c}) \approx {1.1}$meV (assuming all couplings beyond nearest and next-nearest neighbors are negligible), where the factor of $6$ arises because each atom has six nearest and next nearest neighbors. This is somewhat in tension with our PBE+U results, which from Tb.  \ref{tab:couplings} gives  $6(J_1 + J_2 + J_{1c} + J_{2c})\sim \SI{5.4}{\milli\electronvolt}$. Despite the fairly large overestimate, our PBE+U calculations, with $U=\SI{0.3}{\electronvolt}$ so that $D\sim\SI{1}{\milli\electronvolt}$, notably yield reliable \emph{relative} values of exchange constants consistent with the estimates based on our experiments. Our choice of $U$ also leads to The AFM stripy phase being predicted to have the lowest energy of all collinear magnetic configurations examined, in line with the results of our classical model and previous neutron data \cite{Suzuki1993}. Moreover, the tendency for DFT+U to overestimate exchange constants at small or near-zero values of U is well documented \cite{Lee2016, Loschen2007, Martin1997}. Knowing our PBE+U results likely overestimate the Heisenberg couplings while capturing their relative values, following previous work \cite{Linneweber2017} we uniformly scale $J_1$,$J_2$, $J_{1c}$ and $J_{2c}$ so that $6(J_1 + J_2 +  \cdots )= \SI{1.1}{\milli\electronvolt}$, in line with our Curie-Weiss data, and closely agreeing with the data in Ref. \cite{cw4p3_2plus}, whose fitted temperatures predict a slightly higher $\sum_i J_i \approx \SI{1.3}{\milli\electronvolt}$.

Taking the new parameters $(J_1,J_2,J_{1c},J_{2c},D)=(0.15,-0.0012,0.077,-0.044,1.09)$ \si{\milli\electronvolt}, we can semi-quantitatively reproduce the magnetization curve. We estimate the $g$-factor as $g=2.09=g_\text{Fe}$ \cite{singh1976calculation}. This yields estimated critical fields of \SI{15}{\tesla} and \SI{45}{\tesla}, as shown in Fig. \ref{fig:magnetization}. With no fitting to the experimental magnetization in Fig. \ref{fig:magnetization}, we already have found remarkable agreement between theory and experiment. It is shown in the Supplement \cite{SupMat} that changing the $J$ values slightly - within a range still consistent with our expectations based on Curie-Weiss - moves the transition fields into even better agreement. 

The UUUD phase responsible for the half-magnetization plateau is stable at the classical level over a wide range of applied fields. The model Eq. \eqref{eq:mod_ham} qualitatively reproduces the critical field strengths and quantitatively captures the magnitude of the magnetization. However, it fails to describe some of the fine features of the measurements, such as the small, positive slope of the magnetization within plateaus and the intermediate phase detected by measurements between the plateau and stripy order. The symmetry constraints of the switching reported in Ref. \cite{nair_electrical_2019} also indicate an in-plane component to the moment at zero field which is not accounted for in this model. To capture the remaining fine features of \ce{Fe_{1/3}NbS2} would require a more sophisticated 3D model with vastly more parameters and temperature effects, similar to \cite{mankovsky_electronic_2016,seabra2011competition}. On the other hand, as a minimal model that only includes a subset of the degrees of freedom, the model is highly consistent with measurements and seems to have identified the dominant interactions responsible for the magnetization response of \ce{Fe_{1/3}NbS2}.\\

The applicability of the lattice model suggests that \ce{Fe_{1/3}NbS_2} is proximate to many other phases, some of which are possibly similar to supersolid phases discussed by Seabra and Shannon~\cite{seabra2011competition}. One of these may describe the boundary phase dividing stripy and plateau orders in Fig.~\ref{fig:phasediagHT}. As described in the Supplement~\cite{SupMat}, this intermediate phase has a kink in the magnetization at approximately $\sim{}1.6\mu_B$, or around a third of the saturation magnetization. This may associate the intermediate phase with an UUD transition which appears naturally in the same lattice models~\cite{seabra2011competition}, but our data is inconclusive on this matter. 

The agreement of the experimentally observed magnetization with a classical model suggests that the magnetic behavior, while originating from many competing interactions, involves conventional magnetic phases. This model could be further confirmed by inelastic neutron scattering. The existence of an UUUD half-magnetization plateau had previously been studied as a result of strong next-nearest neighbor interactions within the triangular-lattice plane; we have determined that it is not limited to that case, as we see its emergence from strong interplanar interactions. The determination of these interactions and of the abnormally strong single-ion anisotropy has a large impact on the zero-field ground state of this material; the three-fold symmetry breaking seen in optical measurements~\cite{little_observation_2019}, for example, originates from a magnetic order driven by a large ratio of $J_{1c}/J_1\approx 1/2$, likely stripy in nature with a significant $c$-axis component. It is interesting to consider the implications for the electrical switching of the spin texture of this material. In the typical mechanism, an in-plane N\'eel vector can be naturally rotated by the angular momentum imparted by an in-plane spin polarized current. In contrast, \ce{Fe_{1/3}NbS_2} has a N\'eel vector that is predominantly pointed out-of-plane, so that a different kind of mechanism to transfer angular momentum is likely to be active. The present work suggests that this leverages both strong inter- and intra-planar exchange interactions.

\begin{acknowledgments}
This work was supported as part of the Center for Novel Pathways to Quantum Coherence in Materials, an Energy Frontier Research Center funded by the US Department of Energy, Office of Science, Basic Energy Sciences. Work by J.G.A. and S.C.H. was funded in part by the Gordon and Betty Moore Foundation’s EPiQS Initiative, Grant GBMF9067 to J.G.A. Work by T.C. and D.E.P. was supported by NSF  Graduate  Research  Fellowship Program, NSF DGE 1752814. A portion of this work was performed at the National High Magnetic Field Laboratory, which is supported by National Science Foundation Cooperative Agreement No. DMR-1644779 and the State of Florida.
\end{acknowledgments}

%

\end{document}